\documentclass[longbibliography, aps, prx, amsmath, amssymb, amsfonts, twocolumn,superscriptaddress, footinbib]{revtex4-2}
\usepackage[german,american]{babel}
\usepackage{graphicx}
\usepackage{graphics}
\usepackage{dcolumn}
\usepackage{bm}
\usepackage{amssymb}
\usepackage{amsmath}
\usepackage{amsfonts}
\usepackage{epsfig}
\usepackage{mathbbol}
\usepackage{latexsym}
\usepackage{dcolumn}
\usepackage{subfigure}
\usepackage{hyperref}
\usepackage{float}
\usepackage[normalem]{ulem}
\usepackage[usenames, dvipsnames]{color}
\usepackage{epstopdf}
\usepackage{tikz}
\usetikzlibrary{quantikz2}
\usepackage{verbatim}  
\usepackage{mdframed}
\usepackage{color}
\usepackage{algorithm}
\usepackage{algorithmic}

\usepackage{theorem}
\newtheorem{theorem}{Theorem}
\newtheorem{definition}[theorem]{Definition}

\newtheorem{conjecture}{Conjecture}

\newcommand\titlelowercase[1]{\texorpdfstring{\lowercase{#1}}{#1}}

\hypersetup{                    
	colorlinks,
	citecolor=blue,
	filecolor=blue,
	linkcolor=blue,
	urlcolor=blue
}
\usepackage[utf8]{inputenc}

\begin{document}
	\title{Variational learning algorithms for quantum query complexity}
    \author{Zipeng Wu}
    \affiliation{Department of Physics, The Hong Kong University of Science and Technology,\\ Clear Water Bay, Kowloon, Hong Kong, China}
    \author{Shi-Yao Hou}
    \affiliation{College of Physics and Electronic Engineering, Center for Computational Sciences,  Sichuan Normal University, Chengdu 610068, China}
    \affiliation{Shenzhen SpinQ Technology Co., Ltd., Shenzhen, China}
    \author{Chao Zhang}
    \affiliation{Department of Physics, The Hong Kong University of Science and Technology,\\ Clear Water Bay, Kowloon, Hong Kong, China}
    \author{Lvzhou Li}
    \affiliation{Institute of Quantum Computing and Computer Theory, School of Computer
Science and Engineering, Sun Yat-sen University, Guangzhou 510006, China}
	\author{Bei Zeng}
	\email{zengb@ust.hk}
	\affiliation{Department of Physics, The Hong Kong University of Science and Technology,\\ Clear Water Bay, Kowloon, Hong Kong, China}

	\begin{abstract}
		Quantum query complexity is pivotal in the analysis of quantum algorithms, encompassing well-known examples like search and period-finding algorithms. These algorithms typically involve a sequence of unitary 
		operations and oracle calls dependent on an input variable. In this study, we introduce a variational learning approach 
		to explore quantum query complexity. Our method employs an efficient parameterization of the unitary operations and 
		utilizes a loss function derived from the algorithm's error probability. We apply this technique to various quantum 
		query complexities, notably devising a new algorithm that resolves the 5-bit Hamming modulo problem with four queries, 
		addressing an open question from arXiv:2112.14682. This finding is corroborated by a Semidefinite Programming (SDP) 
		approach. Our numerical method exhibits superior memory efficiency compared to SDP and can identify quantum query 
		algorithms that require a smaller workspace register dimension, an aspect not optimized by SDP. These advancements 
		present a significant step forward in the practical application and understanding of quantum query algorithms.
	\end{abstract}
	\date{\today}
	\maketitle
	
	\section{Introduction}\label{Introduction}

    Query models (also known as the decision tree model~\cite{buhrman2002complexity}) play important roles in analyzing 
	quantum algorithms. It captures most of the known quantum algorithms, such as search ~\cite{grover1996fast}, 
	period-finding~\cite{shor1999polynomial}, Simon's algorithm \cite{Simon1997On}, and the Deutsch-Jozsa 
	algorithm \cite{deutsch1992rapid}. It also can be used as a tool to analyze lower bounds for quantum 
	algorithms~\cite{beals2001quantum, ambainis2002quantum,ambainis2018understanding}. There has been a 
	large body of references, e.g., \cite{ambainis2014exact,ambainis2016superlinear,aaronson2016separations,ambainis2017separations,aaronson2018forrelation,tal2020towards,aaronson2016polynomials,arunachalam2019quantum,chen2020characterization,ye2020characterization},  exploring the advantage of quantum  computing relative to classical computing in the query  model. It has been shown that exponential separations between quantum and classical query complexity can be obtained for computing partial Boolean functions, whereas quantum query algorithms can only achieve polynomial speed-up over classical counterparts for computing total Boolean functions \cite{beals2001quantum}.
	
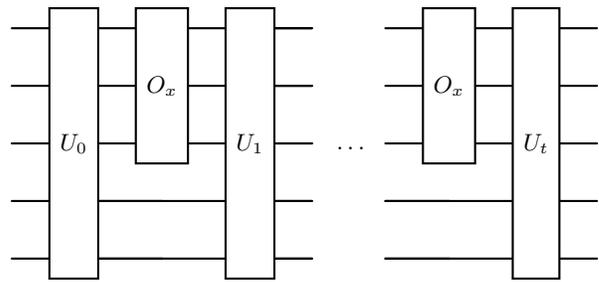
\begin{figure}

    	\begin{quantikz}
		& \gate[5]{\textsc{$U_0$}} & \gate[3]{\textsc{$O_x$}}
		 &\gate[5]{\textsc{$U_1$}} & \qw \\
		& 	&	&  &   \qw \\
		&	&	&  &   \qw \\
		&	&\qw	&  &   \qw \\
		&	&\qw	&  &   \qw 
	\end{quantikz}
	\ $\dots$
	\begin{quantikz}
		    & \gate[3]{\textsc{$O_x$}} & \gate[5]{\textsc{$U_t$}} &   \\
		 	&						   &  						  &   \\
			&						   &  						  &   \\
			&\qw	                   &  		                  &   \\
			&\qw	                   &  		                  &   
	\end{quantikz}
        \centering
        \caption{The circuit structure of $t$-step quantum query algorithm. Oracle acts on the query register, and unitary acts on the whole accessible space. }    
        \label{fig:QQA-structure}
    \end{figure}

	
    A quantum query algorithm (QQA) is defined by an initial state, which, without loss of generality, 
	can be chosen as the all-zero state $|0\rangle^{n}$, and transformations $U_0O_xU_1\ldots O_x U_t$~\cite{ambainis2018understanding}. 
	Here $O_x$ is the oracle that depends on some input variable $x$, and $U_i$s are unitary operations independent of $x$. 
	The algorithm is to apply $U_tO_x\cdots U_1O_xU_0$ on the input state $|0\rangle^{n}$ and measure the result, as shown in 
	FIG.\ref{fig:QQA-structure}.  The quantum query complexities in bounded-error and exact settings
	 are denoted by \(Q_{\varepsilon}(f)\) and \(Q_E(f)\) for Boolean function $f$, respectively. 
	In the bounded-error setting, \(Q_{\varepsilon}(f)\) represents the minimum number of queries required to solve a 
	problem with a bounded probability of error, while in the exact setting, \(Q_E(f)\) signifies the 
	minimum number of queries needed to ascertain a solution with zero-error.

	The intricacies of determining the quantum query complexity of a particular function \(f\) have beckoned
	 the development of distinct methods. Two notable methodologies are the polynomial method\cite{beals2001quantum} 
	 and the adversary \cite{ambainis2000quantum} method. The former hinges on representing or approximating an algorithm computing a Boolean function \(f\) 
	 with a real-valued polynomial, while the latter underscores the limited information gleaned from oracle calls 
	 representing the function \(f\). These methods have proven potent in characterizing bounds for quantum query 
	 complexities in both bounded-error and exact settings.

    The work of Barnum et al.\cite{barnum2003quantum} notably offers a precise determination of \(Q_E(f)\) and \(Q_{\varepsilon}(f)\)  through a 
	Semidefinite Programming (SDP) characterization of the quantum query complexity. This method, being a 
	variant of the adversary method, has paved the way for a numerical approach to analyze and construct 
	quantum query algorithms. This numerical avenue was further explored by Montanaro et al.  \cite{montanaro2015exact}, who, inspired 
	by numerical results, designed an extended form of the Deutsch-Jozsa algorithm and provided numerical 
	results on the optimal success probabilities of quantum algorithms computing all boolean functions on 
	up to 4 bits, and all symmetric boolean functions on up to 6 bits. 

	Despite these advancements, the realm of exact quantum query complexity continues to harbor numerous 
	open questions. For instance, the exact quantum query complexity \(Q_E(f)\) for many Boolean functions, whether
	 symmetric or asymmetric, remains elusive. Furthermore, the quantum speed-up factor had been stagnant at a mere
	  factor of 2 for many years \cite{cleve1998quantum}, until a breakthrough by Ambainis\cite{ambainis2016superlinear} introduced a total Boolean function with a 
	  superlinear advantage of exact quantum algorithms over their classical counterparts. The pursuit of total 
	  Boolean functions with acceleration factors surpassing 2 remains a significant challenge in the field 
	  of exact quantum query complexity.

The SDP method, while powerful, grapples with limitations when addressing these open questions, especially for total Boolean functions where the exponentially growing number of optimization parameters renders the SDP method infeasible for larger-scale Boolean functions. To surmount this challenge, in this paper, we propose a variational heuristic called VarQQA that transmutes the SDP problem into an unconstrained optimization problem, thereby formulating a loss function. This heuristic approach enables the loss function to attain an optimal value of zero when the number of queries \(t\) is at least \(Q_E(f)\), facilitating the delineation of exact quantum query algorithms as heuristic solutions. VarQQA also addresses the redundancy of the workspace register dimension in the SDP method by introducing it as a tunable hyperparameter in the optimization problem. This adjustment allows for the reduction of the workspace register dimension post determination of \(Q_E(f)\), thereby optimizing the quantum query algorithm further through heuristic refinements.
   
		To demonstrate the advantages of our method, in this paper, we employ our VarQQA method to compute \(Q_E(f)\) 
		for two Boolean functions. The first one is the Hamming weight modulo function $\mathrm{MOD_{m}^n}$. In the paper by 
		Cornelissen et al.\cite{cornelissen2021exact}, an open problem regarding this function was proposed, 
		further conjecturing that \(Q_E(\mathrm{MOD}_p^p) = p-1\) 
		 for all prime numbers \(p\). In the case of \(p=5\), our method affirmed that its exact query complexity 
		 is 4. Additionally, our algorithm discovered that only one qubit is required as a workspace register. 
		 We postulate that this is the optimal quantum query algorithm for computing \(\mathrm{MOD}_5^5\) since it 
		 necessitates the fewest number of qubits and queries. Moreover, we numerically verified the cases for
		  \(p=7\) and \(p=11\), thus lending credence to Cornelissen's conjecture. We also observed that the dimension
		   of the workspace register exponentially grows with \(p\), providing numerical insight for the analytical
		    construction of such quantum query algorithms.

		The other problem we addressed is the \(\mathrm{EXACT}_{k,l}^n\) problem.
		Ambainis et al.\cite{ambainis2017exact}
		provided bounds for $Q_E(\mathrm{EXACT}_{k,l}^n)$ and determined  values for some specific cases. However, 
		for many instances, the exact value of \(Q_E(\mathrm{EXACT}_{k,l}^n)\) within the known upper and lower 
		bounds remains undetermined. We employed our method to compute \(Q_E(\mathrm{EXACT}_{k,l}^n)\) values for some of the instances 
		not previously determined. Notably, in some of the cases, we calculated the scenario for \(n=16\), which 
		far exceeds the computational reach of the SDP method.

		Our method presents a promising alternative numerical tool for studying quantum query complexity
		. The remainder of this paper is organized as follows: In Sec.~\ref{Sec:Quantum Query Algorithm}, 
		we review the quantum query algorithm and the SDP method. In Sec.~\ref{Sec:algorithm}, we introduce
		 our variational learning algorithm. In Sec.~\ref{Sec:results}, we apply our method to study several 
		 cases of quantum query complexity. We present our conclusions in Sec.~\ref{Conclusion}.

\section{Preliminaries}\label{Sec:Quantum Query Algorithm}

\subsection{The quantum query model}
\label{sec:quantum_query_model}
In a query complexity model (classical or quantum), we wish to compute a function $f: S\to T $ where::
\begin{itemize}
\item $S \subseteq \Sigma^n$ with $\Sigma$ is finite and $n$ is a positive integer.
\item $T$ is a finite set.
\end{itemize}
The input domain consists of points $x = (x_1,...,x_n)$, where $x_i \in \Sigma$, and the output domain is $T$ with cardinality $|T|$. $f$ is a \textit{decision function} if $|T|=2$. $f$ is called a \textit{Boolean function} if $\Sigma=\{0,1\}$, and $f$ is \textit{total} if $S = \Sigma^n$.

Consider a Quantum Query Algorithm (QQA) to compute a Boolean function $f:S\subseteq \{0,1\}^n \to T$. The QQA requires:
\begin{itemize}
\item the \textit{input register} to hold the input $x\in \{0,1\}^n$.
\item the \textit{query register} to hold an integer between $0$ and $n$.
\item the \textit{workspace register} that has no restriction on dimension.
\end{itemize}

The total Hilbert space can be expressed as $H = H_{\text{input}} \otimes H_{\text{Q}} \otimes H_{\text{W}}$. 
The tensor product space of query register and workspace register $H_A=H_{\text{Q}}\otimes H_{\text{W}}$ 
is called the \textit{accessible space}. The dimensions of $H_{\text{input}}$, $H_{\text{Q}}$, and $H_{\text{W}}$ 
are $d_{input}$, $d_q$, and $d_w$ respectively, where $d_{input} = 2^n$, $d_q=n+1$, and $d_w\geq 1$. 
A quantum state in the Hilbert space $H$ can be represented as:
\begin{equation}
|\psi\rangle = \sum_{x,i,w} \alpha_{x,i,w} |x\rangle |i\rangle |w\rangle \label{eq1}
\end{equation}

To compute $f(x)$, the QQA works as follows:
\begin{itemize}
\item[a.] Input the quantum state $|x\rangle |0\rangle |0\rangle$.
\item[b.] Sequentially apply the unitaries $U_i$ that only act on the accessible space and the oracle $O$.
\item[c.] Measure the qubits in the accessible space.
\end{itemize}
The complexity of a quantum query algorithm is determined by the number of queries it calls.

Here, the oracle $O$ satisfies
\begin{equation}
O|x\rangle|i\rangle |w\rangle = (-1)^{x_i}|x\rangle|i\rangle|w\rangle = |x\rangle(-1)^{x_i}|i\rangle|w\rangle
\end{equation}

Since the oracle $O$ leaves the state in the input Hilbert space $H_{\text{input}}$ unchanged, an equivalent QQA works only on the accessible space, where the query operator $O$ is now dependent on the input $x$ and is given by:
\begin{equation}
O_x|i\rangle |w\rangle = (-1)^{x_i}|i\rangle|w\rangle
\label{eqn:oracle_x}
\end{equation}
where $x_0$ is not part of the input and is defined to be the constant $0$. Note that, as mentioned in \cite{barnum2003quantum}, the index $i=0$ is needed in the model for an important technical reason without which the model is not even capable of computing some simple functions. However, in many cases, we do not need the index $i=0$, like in Grover's algorithm. Then, the initial state can be set with $\alpha_{x,i,w}=0$ for $i=0$ in Eq.(\ref{eq1}), and the unitary operations are constructed to act trivially on the corresponding dimension.

The QQA algorithm is then given by:
\begin{itemize}
\item[a.] Input the initial state $|0\rangle|0\rangle$ in accessible space $H_A$.
\item[b.] Sequentially apply the unitaries $U_i$, which are independent of $x$, and the oracle $O_x$ for a specific $x$.
\item[c.] Measure the qubits in the accessible space.
\end{itemize}

A \( t \)-step quantum query algorithm operating in the accessible space is depicted in Figure \ref{fig:QQA-structure}. 
Consider \( |\Psi_x^{(j)}\rangle \) as the quantum state post-application of \( U_j \) for a given 
input \( x \), which can be expressed as
\begin{equation}
|\Psi_x^{(j)}\rangle = U_jO_x...O_xU_0|0\rangle|0\rangle \label{eq2}
\end{equation}
This state can be uniquely decomposed as:
\begin{align}
\label{eq:state}
|\Psi_x^{(j)}\rangle &= \sum_i |i\rangle |\Psi_{x,i}^{(j)}\rangle \\
&\text{where } |i\rangle \in H_Q, \ \ |\Psi_{x,i}^{(j)}\rangle \in H_W. \nonumber 
\end{align}
It's important to note that while \(| \Psi_{x,i}^{(j)} \rangle\) is a vector in \( H_W \), it may not be normalized.

In the quantum query algorithm, results are extracted by performing a projective measurement on the output state.
 This measurement is defined using a Complete Set of Orthogonal Projectors (CSOP). Specifically, the CSOP for the algorithm 
 is represented by the indexed set $\{\Pi_{f(x)}: f(x) \in T \}$ of projectors. These projectors are pairwise 
 orthogonal and satisfy the relation \( \sum_{f(x)\in T}\Pi_{f(x)} = I_A \), where \( I_A \) denotes the identity operator 
 on the accessible space. Given a direct sum decomposition of the accessible space as \( \oplus_{f(x) \in T} H_{f(x)} \) into 
 orthogonal subspaces, there exists a unique CSOP \( \left(\Pi_{f(x)}: f(x) \in T\right) \). In this context, \( \Pi_{f(x)} \) 
 serves as the identity on \( H_{f(x)} \) and zeroes out \( H_{f(x^{\prime})} \) for all \( f(x^{\prime}) \neq f(x) \).

In the QQA, the output for a given input \( x \) is determined by the 
probability \( \langle \Psi_x^{(t)}|\Pi_{f(x)}|\Psi_x^{(t)}\rangle \). This probability reflects the 
likelihood of obtaining the correct output \( f(x) \) upon measurement of the output quantum state.

A QQA is termed \textit{exact} when it always produces the correct output for every input, mathematically expressed as:
\begin{equation}
	\label{eqn:exact}
\langle \Psi_x^{(t)}|\Pi_{f(x)}|\Psi_x^{(t)}\rangle = 1, \ \ \ \forall x \in S
\end{equation}

However, in practical scenarios, a QQA might not always be exact. It is said to compute the function \( f \) with an error margin \( \varepsilon \) if the probability of obtaining the correct output is at least \( 1-\varepsilon \) for every input, as given by:
\begin{equation}\label{eqn:success rate}
\langle \Psi_x^{(t)}|\Pi_{f(x)}|\Psi_x^{(t)}\rangle \geq 1-\varepsilon, \ \ \ \forall x \in S
\end{equation}
This criterion allows for a certain degree of tolerance in the algorithm's accuracy while still being considered effective.

The \textit{exact quantum query complexity}, \( Q_E(f) \), is the fewest queries any quantum 
algorithm needs to compute \( f(x) \) accurately for all \( x \). Similarly, 
the \textit{bounded-error quantum query complexity}, \( Q_{\varepsilon}(f) \), represents the 
minimal queries needed to compute \( f \) with an error chance of $\varepsilon$.

\subsection{Semidefinite Programming Formulation for Quantum Query Algorithms}

In this section, we revisit the method of Barnum et al.\cite{barnum2003quantum}, specifically addressing how quantum query algorithms can be represented as semidefinite programs.

We begin by introducing the concept of the Gram matrix. Consider an indexed family of vectors \( \left(|\Psi_x\rangle : x \in S\right) \) within the Hilbert space \( H \). The associated Gram matrix, denoted as \( \operatorname{Gram}\left(|\Psi_x\rangle \right.$ : \( x \in S \) ), is an \( |S| \times |S| \) matrix \( M \) defined by:
\[
M[x, y] \triangleq \left\langle\Psi_x \mid \Psi_y\right\rangle .
\]

We now define the Gram matrix during the QQA. The states defined in Eqn.(\ref{eq2}) and (\ref{eq:state}) generate a sequence that is interlinked by the application of two successive unitaries: the oracle \( O_x \) corresponding to the input \( x \), and the input-independent unitary operator \( U_j \). The final step of the algorithm (measurement) is specified by the orthonormal projections \( \Pi_z \), each corresponding to a different output \( z \in T \).

We define the following symmetric matrices \( M^{(j)}, M_i^{(j)} \), and \( \Gamma_z \) with the matrix elements
\begin{align}
	M^{(j)}[x,y] & \triangleq\left\langle\Psi_x^{(j)} \mid \Psi_y^{(j)}\right\rangle, \\
	M_i^{(j)}[x,y] & \triangleq\left\langle\Psi_{x, i}^{(j)} \mid \Psi_{y, i}^{(j)}\right\rangle, \\
	\Gamma_z[x,y] & \triangleq\left\langle\Psi_x^{(j)} \Pi_z^{\dagger} \mid \Pi_z \Psi_y^{(j)}\right\rangle .
\end{align}

With the symmetric matrices defined, we are now equipped to present the SDP formulation for the quantum query algorithm.

\begin{mdframed}
\begin{definition} [SDP Formulation for QQA]
	Given a Boolean function \( f \), and a number of queries \( t \), solving the following SDP problem yielding the optimal value \( \varepsilon^{\star} \) is equivalent to finding a \( t \)-step QQA that computes \( f \) within error \( \varepsilon^{\star} \).
\end{definition}

\begin{eqnarray}
\textrm{minimize} && \varepsilon \\
\textrm{s.t.} \quad
\label{eqn:init}
\sum_{i=0}^{n} M_{i}^{(0)} &=& E_0 \\
\label{eqn:relation}
\sum_{i=0}^{n} M_{i}^{(j)} &=& \sum_{i=0}^{n} E_{i} \odot M_{i}^{(j-1)} \\
&& \mathrm{for} \ 1 \leq j \leq t-1 \nonumber \\
\label{eqn:relation2}
M^{(t)} &=& \sum_{i=0}^{n} E_{i} \odot M_{i}^{(t-1)} \\
\label{eqn:complete}
M^{(t)} &=& \sum_{z \in T} \Gamma_{z} \\
\label{eqn:error}
\Delta_{z} \odot \Gamma_{z} &\geq& (1-\varepsilon) \Delta_{z} \ \ \mathrm{for} \ z \in T
\end{eqnarray}
where \( \odot \) denotes the element-wise product, \( E_0 \) is the constant 1 matrix, \( E_i[x,y] = (-1)^{x_i+y_i} \) for \( 0 < i \leq n \). \( \Delta_z \) is a diagonal matrix with entries \( \Delta_z[x,x] = 1 \) if \( f(x) = z \) and \( \Delta_z[x,x] = 0 \) otherwise. 
\end{mdframed}
\begin{theorem}[\textnormal{Barnum, Saks, and Szegedy \cite{barnum2003quantum}}]
\ \ \  \ \ \  \ \ \ \\
Let $f : S \subseteq \{0, 1\}^n \to T$ be a Boolean function. There exists a $t$-query QQA($f$) that computes function $f$ within error $\varepsilon$ if and only if SDP($f,t,\varepsilon$)
with conditions (\ref{eqn:init}),(\ref{eqn:relation}),(\ref{eqn:relation2}),(\ref{eqn:complete}) and (\ref{eqn:error}) is feasible. Furthermore, 
let \( r = \max(\mathrm{rank}(M^{(j)}_{i})) \), the necessary dimension of the workspace register is $r$.
\end{theorem}

Solving the SDP(\( f,t,\varepsilon \)) to obtain the optimal value \( \varepsilon^{\star} \) demonstrates that a quantum algorithm with \( t \) queries can compute the function \( f \) with an error rate of \( \varepsilon^{\star} \). Therefore, determining the exact quantum query complexity of the function \( f \) translates to identifying the minimum number of queries, \( t_{min} \), for which the optimal value \( \varepsilon^{\star} \) of SDP(\( f,t_{min},\varepsilon \)) is zero. Montanaro et al.~\cite{montanaro2015exact} considered an \( \varepsilon^{\star} \) value less than 0.001 as indicative of the existence of an exact quantum query algorithm. Thus, the goal of computing \( Q_E(f) \) with SDP is to determine the smallest \( t \) where \( \varepsilon \) falls below this threshold.

SDP benefits from the robust framework of convex optimization but encounters significant 
challenges in practical applications. A critical issue is the exponential growth in the 
number of decision variables, which correspond to the optimization parameters. This growth 
is directly linked to the input set cardinality for Boolean functions, presenting difficulties 
in solving even moderately sized problems. Specifically, the memory requirements for SDP(\(f,t,\varepsilon\)), primarily due to \(M^{(j)}_i\) as indicated in Eq.\ref{eqn:relation}, scale as \(O(n t|S|^2)\). With \(|S| = 2^n\) for total Boolean functions, the memory demand escalates to \(O(t n  4^n)\). Our numerical tests show that this scaling renders SDP(\(f,t,\varepsilon\)) inefficient for Boolean functions with an input length of \(n=10\), even on a server equipped with 256GB of memory.

Furthermore, the SDP method does not optimize the dimension of the workspace register, \(d_w\), in the computation of \(Q_E(f)\). This often results in a situation where \( \max(\text{rank}(M_i^{(j)})) \) is approximately equal to \(|S|\), suggesting that the derived quantum query algorithms (QQAs) typically require the maximum possible \(d_w\). However, our investigations using the VarQQA approach indicate that the actual \(d_w\) necessary for these computations is frequently lower than the maximum predicted by SDP.

\section{The V\titlelowercase{ar}QQA Method}\label{Sec:algorithm}

In addressing quantum query complexity, we pivot from the traditional Semidefinite Programming (SDP) approach to a variational learning strategy, commonly employed in machine learning (ML) for tackling complex optimization challenges. As illustrated in FIG.\ref{fig:QQA-structure}, we parameterize the unitaries using free parameters, transforming the constrained convex optimization of SDP into an unconstrained landscape. Transitioning to an unconstrained optimization framework not only aligns with established ML optimization strategies but also opens the door to a plethora of advanced techniques tailored for such problems. Specifically, we can capitalize on the power of auto-differentiation for efficient gradient computations, a hallmark of modern ML methodologies. Additionally, for the optimization itself, we employ the Limited-memory BFGS method\cite{liu1989limited}, renowned for its efficiency in handling large-scale problems. \
It's important to clarify that the VarQQA framework operates as a heuristic, leveraging these advanced ML techniques to explore quantum query challenges in a 
more adept and streamlined manner, rather than functioning strictly as an algorithm with 
guaranteed optimal solutions.
\subsection{Formulation of VarQQA}
Given a Boolean function and a specified number of queries \( t \), our objective is to discern the optimal quantum query algorithm. Here, by ``algorithm," we specifically refer to the configuration of each unitary in the circuit. The term ``optimal" is used in the context of achieving the minimal value of the loss function. VarQQA is designed to address this challenge. It parameterizes the unitary operations with trainable parameters and crafts a loss function intrinsically linked to the error rate. Through iterative refinement of these parameters, VarQQA seeks to minimize the loss function, thereby pinpointing the quantum query algorithm configuration that yields the lowest error for the Boolean function \( f \) with \( t \) queries.

Central to our VarQQA method is the design and implementation of two foundational components: the parameterization of unitaries and the formulation of the loss function. For the unitary parameterization, we employ a direct exponential of Hermitian matrices, ensuring a compact and efficient representation of our quantum operations. A detailed discussion of this parameterization technique is provided in Appendix \ref{appendix:param}. Transitioning to the loss function, it stands as a pivotal element in our variational framework, offering a quantifiable metric to gauge the accuracy and efficacy of our quantum queries. Let's delve deeper into the specifics of this loss function.

\begin{definition}[Loss Function for VarQQA]
	Consider a \( t \)-step quantum query algorithm where each unitary operation is parameterized as \( U_0(\theta_0), \dots, U_t(\theta_t) \). The output state of the algorithm for an input \( x \) is represented by:
	\[
	|\Psi_x^{(t)}( \bm{\theta} ) \rangle = U_t(\theta_t)O_xU_{t-1}(\theta_{t-1}) \dots O_xU_0(\theta_0)|0\rangle|0\rangle.
	\]
	Given the complete set of orthogonal projectors \( \{\Pi_{f(x)} \mid f(x) \in T \} \), we introduce the loss function \( \mathcal{E} \) as:
	\begin{equation}
	\mathcal{E} = \frac{1}{|S|} \sum_{x \in S} \langle \Psi_x^{(t)}( \bm{\theta} )| \Pi_{\bot f(x) } |\Psi_x^{(t)}( \bm{\theta} ) \rangle,
	\end{equation}
	where \( \Pi_{\bot f(x)} \) is defined as:
	\begin{equation}
	\Pi_{\bot f(x)} = I_A - \Pi_{f(x) },
	\end{equation}
	and it denotes the projector onto the orthogonal complement of the space associated with \( f(x) \).
\end{definition}

The loss function \( \mathcal{E} \), defined as the average error rate over all inputs \( x \), is inherently non-negative, 
reflecting the nature of error rates. As highlighted by Eq.\ref{eqn:exact}, an ``exact" quantum query algorithm is one where the error rate for each input \( x \) is zero. Therefore, \( \mathcal{E} \) reaches its minimum value of zero if and only if the algorithm yields no error across all inputs. This zero minimum is indicative of the algorithm's exactness, a state achieved only when every individual error rate, contributing to the average, is zero.

We are now prepared to present our VarQQA, as detailed in the learning process outlined in \textbf{Algorithm \ref{alg}}.
\begin{algorithm}[H]
	\renewcommand{\algorithmicrequire}{\textbf{Input:}}
	\renewcommand{\algorithmicensure}{\textbf{Output:}}
	\caption{VarQQA}
	\label{alg}
	\begin{algorithmic}[1]
	\REQUIRE Tolerance $\varepsilon_{tol}$, \\ 
	 \ \ \ \ \ \  Unitary Parametrization $\{U_0(\theta_0),U_1(\theta_1),...,U_t(\theta_t)\}$, \\
     \ \ \ \ \ \  \ \  Complete set of Orthogonal Projectors$\{\Pi_{f(x)}\ | x\in S \}$ \\
	\STATE Initialization: $\bm{\theta} = \{\theta_0,\theta_1,...,\theta_t\}$
	\STATE \textbf{repeat}
	\STATE $|\Psi_x^{(t)}( \bm{\theta} ) \rangle = U_t(\theta_t)O_xU_{t-1}(\theta_{t-1}) \dots O_xU_0(\theta_0)|0\rangle|0\rangle$
	\STATE $\mathcal{E} = \frac{1}{|S|} \sum_{x \in S} \langle \Psi_x^{(t)}( \bm{\theta} )| I_A-\Pi_{f(x)} |\Psi_x^{(t)}( \bm{\theta} ) \rangle$
	\STATE $\nabla \mathcal{E} (\text{gradient backpropagation)}$
	\STATE Update $\{\theta_0,\theta_1,...,\theta_t\}$ based on $\nabla \mathcal{E}$
	\STATE \textbf{until} $\mathcal{E} < \varepsilon_{tol}$ or $\mathcal{E}$ converges
	\ENSURE $ \mathcal{E}, \{U_0(\theta_0),U_1(\theta_1),...,U_t(\theta_t)\}$
	\end{algorithmic}
\end{algorithm}
The learning process of VarQQA can be divided into the forward pass and the backward pass.
The forward pass is quantum circuit evolution that is computed by matrix-vector multiplications.
We implement a vectorized version of the forward pass to speed up the calculation, i.e.,

we forward the unitaries and oracles for all input \( x \) simultaneously. This can greatly harness the speed of a GPU.
In the backward pass, the gradient of parameters \( \theta \) is calculated automatically by the PyTorch framework\cite{paszke2019pytorch}.
With all gradients calculated, we can update the trainable parameters with the limited-memory BFGS algorithm in the Scipy package\cite{2020SciPy-NMeth}.
\subsection{Searching for \( Q_E(f) \) with VarQQA}
Given a Boolean function \( f \), we can apply VarQQA to search for its \( Q_E(f) \). 
The methodology is similar to SDP methods; we gradually increase the number of queries \( t \) until the minimal error rate \( \mathcal{E^{\star}} \) reaches below a threshold \( \epsilon \).
In this work, we choose \( \epsilon = 10^{-5} \) as the threshold. This means that when we find a quantum query algorithm with an error rate \( \mathcal{E} < 10^{-5} \), 
we consider it as an exact quantum query algorithm.

VarQQA does not inherently ensure convergence to the global minimum due to its non-convex nature.
However, we can employ several strategies to enhance the likelihood of reaching the global optimum. 
Utilizing multiple random initializations allows us to explore various regions of the solution space, 
increasing the probability of finding the global minimum. Advanced optimizers, particularly those with 
adaptive learning rates or momentum-based techniques, can significantly refine the search process.
Additionally, the incorporation of domain-specific knowledge can guide the optimization process
more effectively. For example, when the lower bound of \( Q_E(f) \) is known, VarQQA can leverage this information to 
streamline the search, particularly for large-scale problems where SDP is not viable.

By implementing these strategies, we aim to mitigate the limitations inherent to VarQQA and bolster 
the performance of our algorithm. As demonstrated in the results section (see Sec.\ref{Sec:results}),
VarQQA exhibits a strong capability to find the global minimum, with most algorithms converging to 
it within a few random initializations.

\section{Results}\label{Sec:results}

In this section, we investigate the feasibility of employing our VarQQA method to compute the exact quantum query complexity \( Q_E(f) \) for specific Boolean functions \( f \). We select two total functions, \( \text{MOD}_m^n \) and \( \text{EXACT}_{k,l}^n \), whose \( Q_E(f) \) values are currently open problems. By applying VarQQA to these cases and contrasting the results with those obtained via the SDP method, we aim to demonstrate the algorithm's capability and potential advantages in solving such complex problems.

The implementation of VarQQA and numerical results in this section can be accessed at~\footnote{\url{https://github.com/wuzp15/VarQQA}}.

\subsection{Hamming Weight Modulo Function}
A $n$-bit mod $m$ Hamming weight modulo function is defined as:
\begin{equation}
\mathrm{MOD}^{n}_m(x) = |x| \ \mathrm{mod} \ m , \ \forall x\in\{0,1\}^n   
\end{equation}
where $|x|$ denotes the Hamming weight of $x$.

Recently, Cornelissen et al.\cite{cornelissen2021exact} demonstrated that for \( \mathrm{MOD}_m^n \) where $m$'s prime factors are only 2 or 3, the exact quantum query complexity is \( \left\lceil n\left(1-\frac{1}{m}\right)\right\rceil \). Furthermore, they established that the exact quantum query complexity for any \( 1<m \leq n \) is at least this amount. Following these findings, they proposed Conjecture \ref{conj:hamming}, suggesting that this lower bound is indeed tight.

\begin{conjecture}\label{conj:hamming}
The exact quantum query complexity of \( \mathrm{MOD}_m^n \) is \( \left\lceil n\left(1-\frac{1}{m}\right)\right\rceil \).
\end{conjecture}

The recursive proof method used in their work suggests that Conjecture \ref{conj:hamming} could be resolved if we can prove \( Q_E(\mathrm{MOD}_p^p)=p-1 \) for prime numbers $p$. Given that the cases for $p=2$ and $3$ have been resolved, a pressing question emerges: is it possible to construct a 4-query quantum algorithm to exactly compute \( \mathrm{MOD}_5^5 \)?

We employed the VarQQA method to identify a 4-step QQA capable of computing $\mathrm{MOD}_5^5$ exactly. VarQQA successfully found a QQA that can compute $\mathrm{MOD}_5^5$ with an error rate below $10^{-5}$, utilizing a workspace dimension $d_w=2$. Initially, we started with $d_w=1$, but an exhaustive search did not yield a viable solution. Upon increasing the workspace dimension to $d_w=2$, we achieved an accessible dimension of $d_A = 2 \times (n+1) = 12$. The accessible space was then partitioned into orthogonal subspaces with dimensions (2,4,1,1,4). We posit that the QQA discovered through VarQQA for $\mathrm{MOD}_5^5$ is optimal, as it utilizes the minimal number of queries and the least workspace dimension necessary.

To corroborate our results, the SDP method was also applied to seek a 4-step QQA for $\mathrm{MOD}_5^5$. The SDP approach confirmed the feasibility of solving the Hamming weight modulo problem with an error rate $\varepsilon$ below $10^{-5}$. However, in terms of workspace dimension, the QQA given by SDP required a workspace register with dimension 22 since $\max(\mathrm{rank}(M^{(j)}_{i})) = 22$.

\begin{figure}[h]
	\centering
	\includegraphics[width=9cm]{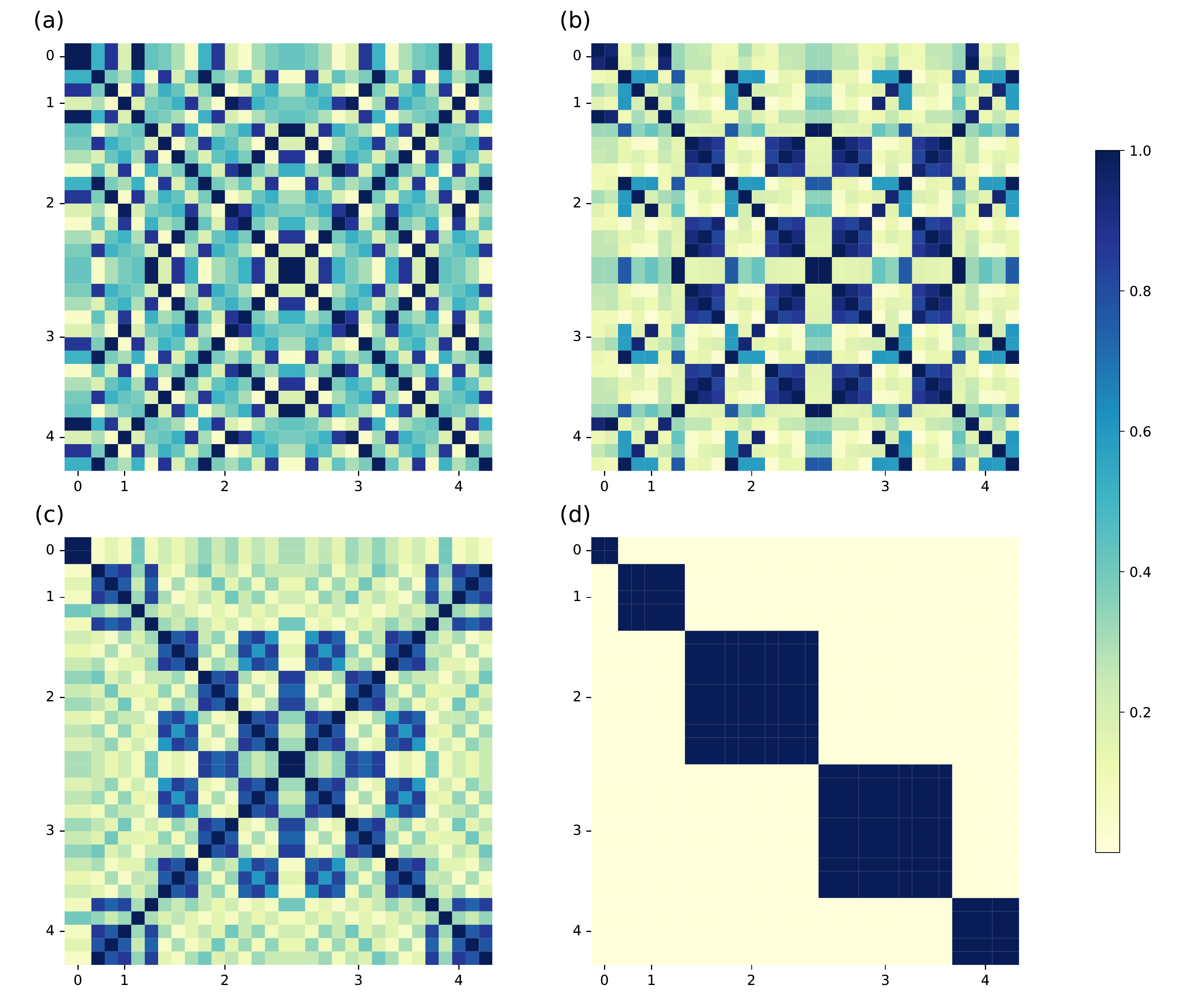}
	\caption{Visualization of Gram Matrices $M^{(j)}$ for quantum states across oracle calls. Subfigures (a), (b), (c), and (d) depict the absolute values of the Gram matrices \( M^{(1)}, M^{(2)}, M^{(3)}, \) and \( M^{(4)} \), respectively. Given the input domain for \( \mathrm{MOD}^5_5 \) is \( \{0,1\}^5 \), each Gram matrix is a \( 32 \times 32 \) representation. The heatmaps capture the magnitude of the inner products \( | \langle \psi^{(i)}_x | \psi^{(i)}_y \rangle| \) for \( i=1,2,3,4 \), indicating the degree of quantum state overlap after 1, 2, 3, and 4 oracle calls. The inputs \( x \) and \( y \) are organized in the heatmaps according to their Hamming weight modulo 5, forming five distinct clusters. Within each cluster, the inputs are sorted by increasing binary value. Notably, as illustrated in subfigure (d), \( |\langle \psi^{(4)}_x | \psi^{(4)}_y \rangle| \) equals 1 if \( x \) and \( y \) share the same Hamming weight modulo value, signifying identical output states modulo a global phase, and 0 otherwise, indicating perfect orthogonality for inputs with different Hamming weight modulo values.}
\label{fig:5bit-combine}
\end{figure}

Further analysis of the Gram Matrix of the output state, denoted as $M^{(4)}[x,y]= \langle \psi_x^{(4)} | \psi_y^{(4)} \rangle$, sheds light on the characteristics of the QQA's output states. It was observed that inputs sharing the same Hamming weight modulo value yielded identical output states, differing only by a global phase, i.e., $\langle \psi_x^{(4)}\mid |\psi_y^{(4)} \rangle = \delta_{\mathrm{MOD}_5^5(x),\mathrm{MOD}_5^5(y)}$. This insight could offer valuable direction for the development of analytical algorithms. The evolution of the Gram matrix after each oracle call is illustrated in FIG.\ref{fig:5bit-combine}.

In a further exploration,we applied VarQQA to search for a 4-step QQA that can ``exactly" compute $\mathrm{MOD}_5^5$ and successfully identified such an algorithm with an error rate below $10^{-7}$.

The workspace dimension $d_w =2$ thus the accessible dimension $d_A=2\times(n+1) = 12$. The accessible space is partitioned into orthogonal subspaces with dimensions (2,4,1,1,4). When we initially explored the case with $d_w=1$, our exhaustive search failed to find a solution, suggesting that a workspace dimension of at least $d_w=2$ is necessary.
 We believe the QQA found by VarQQA for $\mathrm{MOD}_5^5$ is optimal, given that it uses the minimal number of queries and minimal workspace dimension. To validate our findings, we also applied the SDP method to search for a 4-step QQA that computes $\mathrm{MOD}_5^5$. The SDP method confirms that a 4-step QQA can solve the Hamming weight modulo problem with an error rate $\varepsilon$ below $10^{-5}$. However, the SDP results only imply that the workspace dimension is less than 22, as indicated by $\max(\mathrm{rank}(M^{(j)}_{i}))=22$.

The Gram matrix analysis during each stage after calling the oracle provides insights into the output state of the QQA. We found that inputs with the same Hamming weight modulo value have the same output state, differing only by a global phase. The progression of the Gram matrix after each oracle call is depicted in FIG.\ref{fig:5bit-combine}, which may
provide guidance for constructing analytical algorithms.

\begin{table}[]
	\begin{tabular}{|c|c|c|c|c|c|}
		\hline
		$p$  & $Q_{\mathrm{SDP}}$ & $Q_{\mathrm{Var}}$ & $r$ & $d_w$ &  $\mathcal{D}_A$ \\
		\hline
		5  & 4 & 4 & 23 &  2 & [2,4,1,1,4] \\
		7  & 6 & 6 & 115 & 6 & [2,4,7,12,12,7,4] \\
		11 & $\star$ & 10 & $\star$ & 61  & [2,6,18,60,120,161,161,120,60,18,6] \\
		\hline
	\end{tabular}	
\caption{\label{tab:hamming}  Evaluation of $Q_E(\mathrm{MOD}_p^p)$ for primes \( p=5,7,11 \), with the conjectured exact complexity being \( p-1 \). The columns $Q_{\mathrm{SDP}}$ and $Q_{\mathrm{Var}}$ indicate the minimum number of queries required to maintain an error rate \( \varepsilon \) below \( 10^{-5} \), as determined by SDP and VarQQA approaches, respectively. The value \( r=\max(\mathrm{rank}(M^{(j)_i})) \) denotes the maximum workspace dimension deduced from SDP outcomes, whereas \( d_w \) specifies the workspace dimension ascertained through VarQQA. Additionally, \( \mathcal{D}_A \) details the dimensions of each orthogonal subspace within the accessible space, in accordance with the complete set of orthogonal projectors.}
\end{table}

The question of whether a 6-step QQA can exactly compute the $\mathrm{MOD}_7^{7}$ Hamming weight modulo function was also addressed.Indeed, VarQQA found a solution, with an error rate of approximately $10^{-7}$. The accessible space was partitioned into subspaces with dimensions of $(2,4,7,12,12,7,4)$ and a workspace dimension of $d_w=6$. The SDP method also identified a 6-step QQA with an error rate $\varepsilon$ below $10^{-5}$, corroborating our findings.

The dimension of the Gram matrix escalates exponentially in relation to $n$, which prevents the SDP method from delving deeper. Applying VarQQA to search for the $\mathrm{MOD}^{11}_{11}$ case yielded a 10-step QQA that calculates $\mathrm{MOD}^{11}_{11}$ with an error rate around $10^{-6}$. The workspace dimension $d_w$ equals 61, and the accessible space is partitioned into dimensions (2, 6, 18, 60, 120, 161, 161, 120, 60, 18, 6) respectively. These results are summarized in Table \ref{tab:hamming}.

We believe that Conjecture \ref{conj:hamming} generally holds true, and we further conjecture that the workspace dimension of a QQA capable of exactly computing $\mathrm{MOD}^{n}_n$ scales as $d_w \sim O(2^n)$.

\subsection{ $\mathrm{EXACT}^n_{k,l}$ function}
Consider the following $n$-bit function with $0\leq k < l \leq n$:
\begin{equation}
\operatorname{EXACT}_{k, l}^n(x)=\left\{\begin{array}{ll}
1, & \text{if } x_1+\ldots+x_n \in\{k, l\} \\
0, & \text{otherwise}
\end{array}\right.
\end{equation}
i.e., the function returns $1$ only when exactly $k$ or $l$ of the bits $x_i$ are $1$.

Ambainis et al. \cite{ambainis2017exact} proved the following results:
\begin{itemize}
    \item If $l-k = 1$, and $n=k+l$, then $Q_E(\mathrm{EXACT}_{k,k+1}^{2k+1}) = k+1$.
    \item If $l-k \in\{2,3\}$, then $Q_E(\mathrm{EXACT}_{k,l}^{n}) = \max\{n-k,l\}-1$.
    \item For all $k,l$: $\max\{n-k,l\}-1 \leq Q_E(\mathrm{EXACT}_{k,l}^{n}) \leq \max\{n-k,l\}+1$.
\end{itemize}

While previous studies have established a lower bound for $Q_E(\mathrm{EXACT}_{k,l}^n)$, the question of whether these bounds are tight remains open for further investigation. We apply VarQQA to study the unfilled gap of $Q_E(\mathrm{EXACT}_{k,l}^n)$. We claim that an exact quantum query algorithm is very likely to exist when the error rate of the algorithm found by VarQQA is below $10^{-5}$. Montanaro et al. \cite{montanaro2015exact} applied SDP methods to solve exact quantum query complexity for all symmetric Boolean functions up to 6 bits. Therefore, we focus on the case for $n$-bit $\mathrm{EXACT}_{k,l}^{n}$ functions where $n\geq7$.

For the cases where $l-k=1$ and $l+k \neq n$, we used VarQQA to search for $Q_E(\mathrm{EXACT}_{k,l}^n)$ up to 13 qubits. The QQAs that we found by VarQQA are listed in Table \ref{tab:l-k=1}. In most cases, $Q_{\mathrm{Var}}$ meets the theoretical lower bound $Q_L=\max\{n-k,l\}-1$. However, exceptions occur when $k$ equals $(n+1)/2$, including $\mathrm{EXACT}_{4,5}^{7}$, $\mathrm{EXACT}_{5,6}^{9}$, and $\mathrm{EXACT}_{6,7}^{11}$. VarQQA could not find an algorithm to exactly compute $\mathrm{EXACT}_{4,5}^7$ with 4 queries, despite exploring several decompositions of accessible space and increasing the workspace dimension to $10$. Similar behavior was observed when searching for a 5-query algorithm for $\mathrm{EXACT}_{5,6}^9$ and a 6-query algorithm for $\mathrm{EXACT}_{6,7}^{11}$. We also examined the result with SDP and found that the minimum error rate for $\mathrm{EXACT}_{4,5}^7$ with 4 queries converged to approximately $0.001$, and for $\mathrm{EXACT}_{5,6}^{9}$ with 5 queries, the error rate converged to around $0.003$. Therefore, according to the results from both SDP and VarQQA, $Q_E(\mathrm{EXACT}_{4,5}^{7})$ and $Q_E(\mathrm{EXACT}_{5,6}^{9})$ should be 5 and 6, respectively, indicating that the lower bound is not tight for these cases.

For $l-k \geq 2$, we studied the case where $n$ is even, and $k,l$ are symmetrically distributed around $n/2$. As depicted in Table \ref{tab:l-k>2}, all of the cases that we have investigated satisfy $Q_{\mathrm{Var}} = Q_L$.

Both cases highlight the efficiency of VarQQA, revealing that the workspace dimension required by QQA is significantly smaller than the estimates provided by SDP. Furthermore, VarQQA offers insights into the scaling of the workspace dimension \( d_w \) relative to \( n \), providing a new angle to comprehend the complexity of QQA beyond just the number of queries \( t \). 
\begin{table}[]
    \begin{tabular}{|c|c|c|c|c|c|c|}
		\hline
		   ($n,k,l$) & $Q_{L}$ & $Q_{\mathrm{SDP}}$ & $Q_{\mathrm{Var}}$ & $r$ &  $d_w$  & $\mathcal{D}_A$ \\
		   \hline
		   (7,6,7) & 6&6&6 &119&2& [15,1]\\
	
		   (9,8,9) & 8&8&8&502 &2&[19,1] \\
	
		   (11,10,11) &10&$\star$ &10& $\star$ &3 &[35,1] \\
	
		   (13,12,13) &12&$\star$ &12& $\star$ &5& [69,1]\\
		\hline
		   (7,5,6) & 5&5&5&91 & 3 & [22,2]\\
	
		   (9,7,8) & 7&7&7&458 &3 &[29,1] \\
	
		   (11,9,10)& 9 &$\star$ &9& $\star$ &4 &[47,1] \\
	
		   (13,11,12) & 11& $\star$ &11&$\star$ & 5 &[69,1] \\
		\hline 
		   (7,4,5) & 4&5&5&92 &5 &[35,5] \\
	
		   (9,6,7) & 6&6&6&354 &9 &[85,5] \\
	
		   (11,8,9) &8& $\star$&8& $\star$ & 13 & [151,5] \\
	
		   (13,10,11) &10& $\star$&10&$\star$ & 15 & [205,5]\\
		\hline
		(9,5,6) & 5&6&6& 346&  15 &[120,30]\\
	
		(11,7,8) & 7&$\star$&7&$\star$ &  30& [330,30] \\
		\hline
		(11,6,7) & 6&$\star$&7& $\star$ & 45 & [380,160]\\
		\hline
		\end{tabular}
    \caption{$Q_E(\mathrm{EXACT}_{k,l}^{n})$ for $l-k=1$.
Here, $Q_L$ represents the theoretical lower bound of the exact quantum query complexity. 
The columns $Q_{\mathrm{SDP}}$ and $Q_{\mathrm{Var}}$ 
indicate the minimum number of queries required to maintain an error rate \( \varepsilon \) 
below \( 10^{-5} \), as determined by SDP and VarQQA approaches, respectively. The 
value \( r=\max(\mathrm{rank}(M^{(j)_i})) \) denotes the maximum workspace dimension deduced 
from SDP outcomes, whereas \( d_w \) specifies the workspace dimension ascertained through VarQQA.
 Additionally, \( \mathcal{D}_A \) details the dimensions of each orthogonal subspace within 
 the accessible space, in accordance with the complete set of orthogonal projectors.
}
    \label{tab:l-k=1}
\end{table}

\begin{table}[]
    \centering
	\begin{tabular}{|c|c|c|c|c|c|c|}
		\hline
		($n,k,l$) & $Q_{L}$ & $Q_{\mathrm{SDP}}$ & $Q_{\mathrm{Var}}$ & $r$ &  $d_w$  & $\mathcal{D}_A$ \\
		\hline
		(8,2,6) & 5&5&5&127 &3& [25,2]\\
	
		(10,2,8) & 7& 7&7&828 &3&[31,2] \\
	
		(12,2,10) &9& $\star$&9& $\star$ &3 &[38,1] \\
	
		(14,2,12) &11& $\star$&11&$\star$ & 4& [59,1]\\
		(16,2,14) &13& $\star$&13& $\star$ &6& [101,1] \\
		\hline
		(10,3,7) & 6&6&6&461 &8 &[80,8] \\
		
		(12,3,9)& 8& $\star$&8&$\star$ &10 &[124,6] \\
	
		(14,3,11) & 10& $\star$&10&$\star$ & 12 &[176,4] \\
	
		(16,3,13) &12& $\star$&12&$\star$ &14&[236,2]\\
		\hline 
		(12,4,8) & 7& $\star$&7& $\star$ & 30 &[350,40] \\
		(14,4,10) & 9& $\star$&9& $\star$& 38 &[565,5] \\
		\hline
	\end{tabular}
	
    \caption{
$Q_E(\mathrm{EXACT}_{k,l}^{n})$ for $l-k>2$, where $n$ is even and $k, l$ are centered around $n/2$. The notation used in this table is consistent with that in Table \ref{tab:l-k=1}.}
    \label{tab:l-k>2}
\end{table}

\subsection{Numerical analysis of algorithm}
The computational resources utilized in this study are detailed as follows: The semidefinite programming (SDP) calculations were performed on a CPU platform, specifically an Intel Xeon Gold 6248R CPU with 256 GB of RAM. We employed the Splitting Conic Solver (SCS) for these computations and cross-verified the results with the Mosek solver. Conversely, the VarQQA computations were executed on an Nvidia V100 GPU, utilizing the "L-BFGS-B" optimization algorithm.

In the realm of SDP, the number of optimization parameters scales as \( O(t \cdot n \cdot |S|^2) \), where \( t \) represents the number of queries, \( n \) is the bit length of the input domain, and \( |S| \) denotes the cardinality of the input domain. For a total Boolean function, \( |S| \) is equal to \( 2^n \), leading to a scaling of the number of parameters as \( O(t \cdot n \cdot 4^n) \). In contrast, VarQQA parameterizes the unitary operations within the accessible space. A unitary in a \( d \)-dimensional Hilbert space requires \( O(d^2) \) parameters for full parameterization. For an \( n \)-bit Boolean function, the dimension of the accessible space is \( d_q \times d_w \), where \( d_q = n+1 \) and \( d_w \) is specific to the Boolean function \( f \). Hence, VarQQA exhibits an optimization parameter scaling of \( O(t \cdot n^2 \cdot d_w^2) \).

If the dimension of the workspace register scales as \( O(\text{poly}(n)) \), it can be posited that VarQQA would necessitate significantly fewer parameters compared to the SDP method. Our investigation indicates that VarQQA already outperforms the SDP method for certain Boolean functions analyzed in this study when simulated on a classical computer. The SDP method demands an extensive amount of time to solve a 10-bit Boolean function and is incapable of addressing an 11-bit Boolean function due to memory constraints. In stark contrast, VarQQA has successfully determined the exact query complexity for \( \mathrm{EXACT}_{k,l}^{n} \) functions, even for cases with up to 16 bits.

While our method does not inherently assure convergence to a global optimum—a distinguishing feature of convex optimization—this limitation can be mitigated through various strategies. Employing multiple random initializations can enhance the likelihood of locating the global minimum by exploring different regions of the solution space. Furthermore, the application of advanced optimizers, such as those with adaptive learning rates or momentum-based methods, can significantly refine the search process. Additionally, heuristic methods or the incorporation of domain-specific insights can steer the optimization more efficaciously. By implementing these strategies, we aim to offset the inherent limitations and amplify the overall efficacy of our algorithm.

\section{Discussion}\label{Conclusion}

In this study, we introduce a Variational Quantum Query Algorithm (VarQQA) designed to investigate the exact quantum query complexities of Boolean functions. This innovative approach optimizes both the number of oracle calls and the workspace dimension requirements, facilitating the efficient construction of quantum query algorithms. We have applied VarQQA to explore the quantum query complexities of notable Boolean functions, including the Hamming weight modulo function and the $\mathrm{EXACT}_{k,l}^{n}$ function.

Our investigation reveals the limitations of the Semidefinite Programming (SDP) approach, primarily its exponential scaling in optimization parameters and significant memory demands, which confine its applicability to smaller problem instances. In contrast, VarQQA demonstrates scalability and resource efficiency by optimizing a reduced number of parameters, thus enabling the analysis of more complex Boolean functions. Our results attest to VarQQA's superior performance over SDP in managing higher-dimensional problems, underscoring its potential to enhance quantum query complexity analysis and advance quantum computing.

The application of VarQQA to the Hamming weight modulo function $\mathrm{MOD_{m}^n}$ not only supports Cornelissen et al.'s\cite{cornelissen2021exact} conjecture for prime numbers but also highlights the algorithm's minimal resource requirement in terms of queries and qubits. This suggests an optimal quantum query strategy for $\mathrm{MOD}_5^5$, and our extension to larger primes furthers our understanding of quantum query complexities for this function class. Moreover, our exploration of the $\mathrm{EXACT}_{k,l}^n$ problem broadens the known limits of quantum query complexities. VarQQA's ability to address instances previously beyond the computational reach of SDP methods illuminates its potential to bridge significant gaps in our understanding.

These findings not only corroborate existing conjectures but also pave new pathways for the analytical construction of quantum algorithms, emphasizing VarQQA's critical role in propelling forward the domain of quantum computing. A vital direction for future research is the identification of Boolean functions with acceleration factors greater than 2, a key question in the realm of exact quantum query complexity.

	{\textit{Note added:}}  Recently, Ye~\cite{ye2023exact} confirmed Conjecture \ref{conj:hamming} by construction. His algorithm computes the $n$-bit mod $n$ Hamming weight modulo function with $n-1$ queries and requires a $2^n$ dimensional workspace, 
 which aligns with our numerical results. Despite exhibiting the same asymptotic 
 behavior in terms of the workspace dimension, the algorithm discovered by
  VarQQA for $\mathrm{MOD}^5_5$, $\mathrm{MOD}^{7}_7$, and $\mathrm{MOD}^{11}_{11}$ necessitates a smaller workspace.
   This reduction in workspace could potentially save on qubit overhead when utilized 
   as a sub-algorithm for larger scale problems.
	
	\acknowledgments
	
	We thank Chenfeng Cao for helpful discussions. Z-P. Wu, C. Zhang and 
	B. Zeng are supported by GRF grant No. 16305121. S-Y.Hou is supported 
	by National Natural Science Foundation of China under Grant No. 12105195. 
	L. Li is supported by the National Natural Science Foundation of China 
	(Grant No. 61772565) and the Guangdong Basic and Applied Basic Research 
	Foundation (Grant No. 2020B1515020050).

\appendix
\section{Parameterization of $\mathcal{U}(d)$} \label{appendix:param}

In the quest to design Quantum Query Algorithms (QQAs) without prior blueprints, the choice of parameterization for the unitary group \( \mathcal{U}(d) \) is pivotal. The goal is to capture the full expressiveness of \( \mathcal{U}(d) \), allowing for any element to be accurately represented. Among the proposed methods are the Cayley transform\cite{Cayley+1846+119+123}, the Weyl group generators \cite{zhou2003quantum}, recursive constructions \cite{de2018simple}, and the direct matrix exponential of a Hermitian matrix.

The Cayley transform, while efficient and simple to implement, tends to gravitate towards local 
minima during optimization. The recursive method, designed to sample unitaries from the Haar measure,
 becomes increasingly complex with scale, potentially leading to the Barren Plateau phenomenon \cite{mcclean2018barren} and 
 hindering trainning process. The Weyl group approach, although 
 intriguing, also falls prey to local minima and lacks the computational swiftness of the direct
  exponential method. The direct matrix exponential method not only delivers in terms of speed
 but also presents a more conducive optimization landscape. In our experiments with QQAs, we often 
 achieve convergence to the global optimum with fewer than five random initializations.
  
 Our methodology employs the direct matrix exponential technique, utilizing Hermitian 
 parameterization to represent unitary matrices with real parameters. 
 This technique is pivotal in optimization tasks that require finding a optimal set of \( t \) independent 
 unitary matrices \( \{U_1, U_2, \ldots, U_t\} \) within the unitary group \( \mathcal{U}(d) \). Each \( U_j \) is parameterized as \( U_j = e^{iH_j} \),
  where \( H_j \) is a \( d \times d \) Hermitian matrix given by:

 \[
 H_j = A_j + A_j^T + i(B_j - B_j^T)
 \]
 
 Here, \( A_j \) and \( B_j \) are real \( d \times d \) matrices, ensuring that \( H_j \) maintains the Hermitian property, characterized by real diagonal elements and complex conjugate pairs off-diagonally. The construction of \( H_j \) via \( A_j \) and \( B_j \) allows for \( d^2 \) independent real parameters for each \( H_j \), aligning with the dimensionality of \( d \times d \) Hermitian matrices. This parameterization is achieved as follows:
 
 \begin{itemize}
 \item \( A_j + A_j^T \) forms a symmetric matrix from a real upper triangular matrix \( A_j \), offering \( \frac{d(d+1)}{2} \) independent parameters.
 \item \( i(B_j - B_j^T) \) creates a skew-symmetric matrix from a real upper triangular matrix \( B_j \) with zeros along its diagonal, contributing \( \frac{d(d-1)}{2} \) independent parameters. 
 \end{itemize}
 
 The optimization process involves simultaneously adjusting the parameters of all \( H_j \) matrices to minimize the objective function. This is typically achieved using gradient-based optimization methods, where the continuous and differentiable nature of the mapping from \( \mathbb{R}^{d^2} \) to \( \mathcal{U}(d) \) plays a crucial role in the effectiveness of these techniques. To eliminate redundancy associated with the global phase, which does not affect our optimization goals, we set \( \mathrm{tr}(A_j) = 0 \) for each \( A_j \). This parameterization strategy not only ensures efficient gradient computation but also facilitates the optimization process within computational frameworks such as PyTorch, streamlining the search for an optimal set of unitary matrices.
 \bibliography{ref} 

\begin{thebibliography}{33}%
\makeatletter
\providecommand \@ifxundefined [1]{%
 \@ifx{#1\undefined}
}%
\providecommand \@ifnum [1]{%
 \ifnum #1\expandafter \@firstoftwo
 \else \expandafter \@secondoftwo
 \fi
}%
\providecommand \@ifx [1]{%
 \ifx #1\expandafter \@firstoftwo
 \else \expandafter \@secondoftwo
 \fi
}%
\providecommand \natexlab [1]{#1}%
\providecommand \enquote  [1]{``#1''}%
\providecommand \bibnamefont  [1]{#1}%
\providecommand \bibfnamefont [1]{#1}%
\providecommand \citenamefont [1]{#1}%
\providecommand \href@noop [0]{\@secondoftwo}%
\providecommand \href [0]{\begingroup \@sanitize@url \@href}%
\providecommand \@href[1]{\@@startlink{#1}\@@href}%
\providecommand \@@href[1]{\endgroup#1\@@endlink}%
\providecommand \@sanitize@url [0]{\catcode `\\12\catcode `\$12\catcode `\&12\catcode `\#12\catcode `\^12\catcode `\_12\catcode `\%12\relax}%
\providecommand \@@startlink[1]{}%
\providecommand \@@endlink[0]{}%
\providecommand \url  [0]{\begingroup\@sanitize@url \@url }%
\providecommand \@url [1]{\endgroup\@href {#1}{\urlprefix }}%
\providecommand \urlprefix  [0]{URL }%
\providecommand \Eprint [0]{\href }%
\providecommand \doibase [0]{https://doi.org/}%
\providecommand \selectlanguage [0]{\@gobble}%
\providecommand \bibinfo  [0]{\@secondoftwo}%
\providecommand \bibfield  [0]{\@secondoftwo}%
\providecommand \translation [1]{[#1]}%
\providecommand \BibitemOpen [0]{}%
\providecommand \bibitemStop [0]{}%
\providecommand \bibitemNoStop [0]{.\EOS\space}%
\providecommand \EOS [0]{\spacefactor3000\relax}%
\providecommand \BibitemShut  [1]{\csname bibitem#1\endcsname}%
\let\auto@bib@innerbib\@empty
\bibitem [{\citenamefont {Buhrman}\ and\ \citenamefont {De~Wolf}(2002)}]{buhrman2002complexity}%
  \BibitemOpen
  \bibfield  {author} {\bibinfo {author} {\bibfnamefont {H.}~\bibnamefont {Buhrman}}\ and\ \bibinfo {author} {\bibfnamefont {R.}~\bibnamefont {De~Wolf}},\ }\bibfield  {title} {\bibinfo {title} {Complexity measures and decision tree complexity: a survey},\ }\href@noop {} {\bibfield  {journal} {\bibinfo  {journal} {Theoretical Computer Science}\ }\textbf {\bibinfo {volume} {288}},\ \bibinfo {pages} {21} (\bibinfo {year} {2002})}\BibitemShut {NoStop}%
\bibitem [{\citenamefont {Grover}(1996)}]{grover1996fast}%
  \BibitemOpen
  \bibfield  {author} {\bibinfo {author} {\bibfnamefont {L.~K.}\ \bibnamefont {Grover}},\ }\bibfield  {title} {\bibinfo {title} {A fast quantum mechanical algorithm for database search},\ }in\ \href@noop {} {\emph {\bibinfo {booktitle} {Proceedings of the twenty-eighth annual ACM symposium on Theory of computing}}}\ (\bibinfo {year} {1996})\ pp.\ \bibinfo {pages} {212--219}\BibitemShut {NoStop}%
\bibitem [{\citenamefont {Shor}(1999)}]{shor1999polynomial}%
  \BibitemOpen
  \bibfield  {author} {\bibinfo {author} {\bibfnamefont {P.~W.}\ \bibnamefont {Shor}},\ }\bibfield  {title} {\bibinfo {title} {Polynomial-time algorithms for prime factorization and discrete logarithms on a quantum computer},\ }\href@noop {} {\bibfield  {journal} {\bibinfo  {journal} {SIAM review}\ }\textbf {\bibinfo {volume} {41}},\ \bibinfo {pages} {303} (\bibinfo {year} {1999})}\BibitemShut {NoStop}%
\bibitem [{\citenamefont {Simon}(1997)}]{Simon1997On}%
  \BibitemOpen
  \bibfield  {author} {\bibinfo {author} {\bibfnamefont {D.~R.}\ \bibnamefont {Simon}},\ }\bibfield  {title} {\bibinfo {title} {On the power of quantum computation},\ }\href@noop {} {\bibfield  {journal} {\bibinfo  {journal} {SIAM Journal on Computing}\ }\textbf {\bibinfo {volume} {26}},\ \bibinfo {pages} {1474} (\bibinfo {year} {1997})}\BibitemShut {NoStop}%
\bibitem [{\citenamefont {Deutsch}\ and\ \citenamefont {Jozsa}(1992)}]{deutsch1992rapid}%
  \BibitemOpen
  \bibfield  {author} {\bibinfo {author} {\bibfnamefont {D.}~\bibnamefont {Deutsch}}\ and\ \bibinfo {author} {\bibfnamefont {R.}~\bibnamefont {Jozsa}},\ }\bibfield  {title} {\bibinfo {title} {Rapid solution of problems by quantum computation},\ }\href@noop {} {\bibfield  {journal} {\bibinfo  {journal} {Proceedings of the Royal Society of London. Series A: Mathematical and Physical Sciences}\ }\textbf {\bibinfo {volume} {439}},\ \bibinfo {pages} {553} (\bibinfo {year} {1992})}\BibitemShut {NoStop}%
\bibitem [{\citenamefont {Beals}\ \emph {et~al.}(2001)\citenamefont {Beals}, \citenamefont {Buhrman}, \citenamefont {Cleve}, \citenamefont {Mosca},\ and\ \citenamefont {De~Wolf}}]{beals2001quantum}%
  \BibitemOpen
  \bibfield  {author} {\bibinfo {author} {\bibfnamefont {R.}~\bibnamefont {Beals}}, \bibinfo {author} {\bibfnamefont {H.}~\bibnamefont {Buhrman}}, \bibinfo {author} {\bibfnamefont {R.}~\bibnamefont {Cleve}}, \bibinfo {author} {\bibfnamefont {M.}~\bibnamefont {Mosca}},\ and\ \bibinfo {author} {\bibfnamefont {R.}~\bibnamefont {De~Wolf}},\ }\bibfield  {title} {\bibinfo {title} {Quantum lower bounds by polynomials},\ }\href@noop {} {\bibfield  {journal} {\bibinfo  {journal} {Journal of the ACM (JACM)}\ }\textbf {\bibinfo {volume} {48}},\ \bibinfo {pages} {778} (\bibinfo {year} {2001})}\BibitemShut {NoStop}%
\bibitem [{\citenamefont {Ambainis}(2002)}]{ambainis2002quantum}%
  \BibitemOpen
  \bibfield  {author} {\bibinfo {author} {\bibfnamefont {A.}~\bibnamefont {Ambainis}},\ }\bibfield  {title} {\bibinfo {title} {Quantum lower bounds by quantum arguments},\ }\href@noop {} {\bibfield  {journal} {\bibinfo  {journal} {Journal of Computer and System Sciences}\ }\textbf {\bibinfo {volume} {64}},\ \bibinfo {pages} {750} (\bibinfo {year} {2002})}\BibitemShut {NoStop}%
\bibitem [{\citenamefont {Ambainis}(2018)}]{ambainis2018understanding}%
  \BibitemOpen
  \bibfield  {author} {\bibinfo {author} {\bibfnamefont {A.}~\bibnamefont {Ambainis}},\ }\bibfield  {title} {\bibinfo {title} {Understanding quantum algorithms via query complexity},\ }in\ \href@noop {} {\emph {\bibinfo {booktitle} {Proceedings of the International Congress of Mathematicians: Rio de Janeiro 2018}}}\ (\bibinfo {organization} {World Scientific},\ \bibinfo {year} {2018})\ pp.\ \bibinfo {pages} {3265--3285}\BibitemShut {NoStop}%
\bibitem [{\citenamefont {Ambainis}\ \emph {et~al.}(2015)\citenamefont {Ambainis}, \citenamefont {Gruska},\ and\ \citenamefont {Zheng}}]{ambainis2014exact}%
  \BibitemOpen
  \bibfield  {author} {\bibinfo {author} {\bibfnamefont {A.}~\bibnamefont {Ambainis}}, \bibinfo {author} {\bibfnamefont {J.}~\bibnamefont {Gruska}},\ and\ \bibinfo {author} {\bibfnamefont {S.}~\bibnamefont {Zheng}},\ }\bibfield  {title} {\bibinfo {title} {Exact quantum algorithms have advantage for almost all boolean functions},\ }\href@noop {} {\bibfield  {journal} {\bibinfo  {journal} {Quantum Information and Computation}\ }\textbf {\bibinfo {volume} {15}},\ \bibinfo {pages} {35} (\bibinfo {year} {2015})}\BibitemShut {NoStop}%
\bibitem [{\citenamefont {Ambainis}(2016)}]{ambainis2016superlinear}%
  \BibitemOpen
  \bibfield  {author} {\bibinfo {author} {\bibfnamefont {A.}~\bibnamefont {Ambainis}},\ }\bibfield  {title} {\bibinfo {title} {Superlinear advantage for exact quantum algorithms},\ }\href@noop {} {\bibfield  {journal} {\bibinfo  {journal} {SIAM Journal on Computing}\ }\textbf {\bibinfo {volume} {45}},\ \bibinfo {pages} {617} (\bibinfo {year} {2016})}\BibitemShut {NoStop}%
\bibitem [{\citenamefont {Aaronson}\ \emph {et~al.}(2016{\natexlab{a}})\citenamefont {Aaronson}, \citenamefont {Ben-David},\ and\ \citenamefont {Kothari}}]{aaronson2016separations}%
  \BibitemOpen
  \bibfield  {author} {\bibinfo {author} {\bibfnamefont {S.}~\bibnamefont {Aaronson}}, \bibinfo {author} {\bibfnamefont {S.}~\bibnamefont {Ben-David}},\ and\ \bibinfo {author} {\bibfnamefont {R.}~\bibnamefont {Kothari}},\ }\bibfield  {title} {\bibinfo {title} {Separations in query complexity using cheat sheets},\ }in\ \href@noop {} {\emph {\bibinfo {booktitle} {Proceedings of the forty-eighth annual ACM symposium on Theory of Computing}}}\ (\bibinfo {year} {2016})\ pp.\ \bibinfo {pages} {863--876}\BibitemShut {NoStop}%
\bibitem [{\citenamefont {Ambainis}\ \emph {et~al.}(2017{\natexlab{a}})\citenamefont {Ambainis}, \citenamefont {Balodis}, \citenamefont {Belovs}, \citenamefont {Lee}, \citenamefont {Santha},\ and\ \citenamefont {Smotrovs}}]{ambainis2017separations}%
  \BibitemOpen
  \bibfield  {author} {\bibinfo {author} {\bibfnamefont {A.}~\bibnamefont {Ambainis}}, \bibinfo {author} {\bibfnamefont {K.}~\bibnamefont {Balodis}}, \bibinfo {author} {\bibfnamefont {A.}~\bibnamefont {Belovs}}, \bibinfo {author} {\bibfnamefont {T.}~\bibnamefont {Lee}}, \bibinfo {author} {\bibfnamefont {M.}~\bibnamefont {Santha}},\ and\ \bibinfo {author} {\bibfnamefont {J.}~\bibnamefont {Smotrovs}},\ }\bibfield  {title} {\bibinfo {title} {Separations in query complexity based on pointer functions},\ }\href@noop {} {\bibfield  {journal} {\bibinfo  {journal} {Journal of the ACM (JACM)}\ }\textbf {\bibinfo {volume} {64}},\ \bibinfo {pages} {1} (\bibinfo {year} {2017}{\natexlab{a}})}\BibitemShut {NoStop}%
\bibitem [{\citenamefont {Aaronson}\ and\ \citenamefont {Ambainis}(2018)}]{aaronson2018forrelation}%
  \BibitemOpen
  \bibfield  {author} {\bibinfo {author} {\bibfnamefont {S.}~\bibnamefont {Aaronson}}\ and\ \bibinfo {author} {\bibfnamefont {A.}~\bibnamefont {Ambainis}},\ }\bibfield  {title} {\bibinfo {title} {Forrelation: A problem that optimally separates quantum from classical computing},\ }\href@noop {} {\bibfield  {journal} {\bibinfo  {journal} {SIAM Journal on Computing}\ }\textbf {\bibinfo {volume} {47}},\ \bibinfo {pages} {982} (\bibinfo {year} {2018})}\BibitemShut {NoStop}%
\bibitem [{\citenamefont {Tal}(2020)}]{tal2020towards}%
  \BibitemOpen
  \bibfield  {author} {\bibinfo {author} {\bibfnamefont {A.}~\bibnamefont {Tal}},\ }\bibfield  {title} {\bibinfo {title} {Towards optimal separations between quantum and randomized query complexities},\ }in\ \href@noop {} {\emph {\bibinfo {booktitle} {2020 IEEE 61st Annual Symposium on Foundations of Computer Science (FOCS)}}}\ (\bibinfo {organization} {IEEE},\ \bibinfo {year} {2020})\ pp.\ \bibinfo {pages} {228--239}\BibitemShut {NoStop}%
\bibitem [{\citenamefont {Aaronson}\ \emph {et~al.}(2016{\natexlab{b}})\citenamefont {Aaronson}, \citenamefont {Ambainis}, \citenamefont {Iraids}, \citenamefont {Kokainis},\ and\ \citenamefont {Smotrovs}}]{aaronson2016polynomials}%
  \BibitemOpen
  \bibfield  {author} {\bibinfo {author} {\bibfnamefont {S.}~\bibnamefont {Aaronson}}, \bibinfo {author} {\bibfnamefont {A.}~\bibnamefont {Ambainis}}, \bibinfo {author} {\bibfnamefont {J.}~\bibnamefont {Iraids}}, \bibinfo {author} {\bibfnamefont {M.}~\bibnamefont {Kokainis}},\ and\ \bibinfo {author} {\bibfnamefont {J.}~\bibnamefont {Smotrovs}},\ }\bibfield  {title} {\bibinfo {title} {Polynomials, quantum query complexity, and grothendieck's inequality},\ }in\ \href@noop {} {\emph {\bibinfo {booktitle} {Proceedings of the 31st Conference on Computational Complexity}}}\ (\bibinfo {year} {2016})\ pp.\ \bibinfo {pages} {1--19}\BibitemShut {NoStop}%
\bibitem [{\citenamefont {Arunachalam}\ \emph {et~al.}(2019)\citenamefont {Arunachalam}, \citenamefont {Bri{\"e}t},\ and\ \citenamefont {Palazuelos}}]{arunachalam2019quantum}%
  \BibitemOpen
  \bibfield  {author} {\bibinfo {author} {\bibfnamefont {S.}~\bibnamefont {Arunachalam}}, \bibinfo {author} {\bibfnamefont {J.}~\bibnamefont {Bri{\"e}t}},\ and\ \bibinfo {author} {\bibfnamefont {C.}~\bibnamefont {Palazuelos}},\ }\bibfield  {title} {\bibinfo {title} {Quantum query algorithms are completely bounded forms},\ }\href@noop {} {\bibfield  {journal} {\bibinfo  {journal} {SIAM Journal on Computing}\ }\textbf {\bibinfo {volume} {48}},\ \bibinfo {pages} {903} (\bibinfo {year} {2019})}\BibitemShut {NoStop}%
\bibitem [{\citenamefont {Chen}\ \emph {et~al.}(2020)\citenamefont {Chen}, \citenamefont {Ye},\ and\ \citenamefont {Li}}]{chen2020characterization}%
  \BibitemOpen
  \bibfield  {author} {\bibinfo {author} {\bibfnamefont {W.}~\bibnamefont {Chen}}, \bibinfo {author} {\bibfnamefont {Z.}~\bibnamefont {Ye}},\ and\ \bibinfo {author} {\bibfnamefont {L.}~\bibnamefont {Li}},\ }\bibfield  {title} {\bibinfo {title} {Characterization of exact one-query quantum algorithms},\ }\href@noop {} {\bibfield  {journal} {\bibinfo  {journal} {Physical Review A}\ }\textbf {\bibinfo {volume} {101}},\ \bibinfo {pages} {022325} (\bibinfo {year} {2020})}\BibitemShut {NoStop}%
\bibitem [{\citenamefont {Ye}\ and\ \citenamefont {Li}(2020)}]{ye2020characterization}%
  \BibitemOpen
  \bibfield  {author} {\bibinfo {author} {\bibfnamefont {Z.}~\bibnamefont {Ye}}\ and\ \bibinfo {author} {\bibfnamefont {L.}~\bibnamefont {Li}},\ }\bibfield  {title} {\bibinfo {title} {Characterization of exact one-query quantum algorithms (ii): for partial functions},\ }\href@noop {} {\bibfield  {journal} {\bibinfo  {journal} {arXiv preprint arXiv:2008.11998}\ } (\bibinfo {year} {2020})}\BibitemShut {NoStop}%
\bibitem [{\citenamefont {Ambainis}(2000)}]{ambainis2000quantum}%
  \BibitemOpen
  \bibfield  {author} {\bibinfo {author} {\bibfnamefont {A.}~\bibnamefont {Ambainis}},\ }\bibfield  {title} {\bibinfo {title} {Quantum lower bounds by quantum arguments},\ }in\ \href@noop {} {\emph {\bibinfo {booktitle} {Proceedings of the thirty-second annual ACM symposium on Theory of computing}}}\ (\bibinfo {year} {2000})\ pp.\ \bibinfo {pages} {636--643}\BibitemShut {NoStop}%
\bibitem [{\citenamefont {Barnum}\ \emph {et~al.}(2003)\citenamefont {Barnum}, \citenamefont {Saks},\ and\ \citenamefont {Szegedy}}]{barnum2003quantum}%
  \BibitemOpen
  \bibfield  {author} {\bibinfo {author} {\bibfnamefont {H.}~\bibnamefont {Barnum}}, \bibinfo {author} {\bibfnamefont {M.}~\bibnamefont {Saks}},\ and\ \bibinfo {author} {\bibfnamefont {M.}~\bibnamefont {Szegedy}},\ }\bibfield  {title} {\bibinfo {title} {Quantum query complexity and semi-definite programming},\ }in\ \href@noop {} {\emph {\bibinfo {booktitle} {18th IEEE Annual Conference on Computational Complexity, 2003. Proceedings.}}}\ (\bibinfo {organization} {IEEE},\ \bibinfo {year} {2003})\ pp.\ \bibinfo {pages} {179--193}\BibitemShut {NoStop}%
\bibitem [{\citenamefont {Montanaro}\ \emph {et~al.}(2015)\citenamefont {Montanaro}, \citenamefont {Jozsa},\ and\ \citenamefont {Mitchison}}]{montanaro2015exact}%
  \BibitemOpen
  \bibfield  {author} {\bibinfo {author} {\bibfnamefont {A.}~\bibnamefont {Montanaro}}, \bibinfo {author} {\bibfnamefont {R.}~\bibnamefont {Jozsa}},\ and\ \bibinfo {author} {\bibfnamefont {G.}~\bibnamefont {Mitchison}},\ }\bibfield  {title} {\bibinfo {title} {On exact quantum query complexity},\ }\href@noop {} {\bibfield  {journal} {\bibinfo  {journal} {Algorithmica}\ }\textbf {\bibinfo {volume} {71}},\ \bibinfo {pages} {775} (\bibinfo {year} {2015})}\BibitemShut {NoStop}%
\bibitem [{\citenamefont {Cleve}\ \emph {et~al.}(1998)\citenamefont {Cleve}, \citenamefont {Ekert}, \citenamefont {Macchiavello},\ and\ \citenamefont {Mosca}}]{cleve1998quantum}%
  \BibitemOpen
  \bibfield  {author} {\bibinfo {author} {\bibfnamefont {R.}~\bibnamefont {Cleve}}, \bibinfo {author} {\bibfnamefont {A.}~\bibnamefont {Ekert}}, \bibinfo {author} {\bibfnamefont {C.}~\bibnamefont {Macchiavello}},\ and\ \bibinfo {author} {\bibfnamefont {M.}~\bibnamefont {Mosca}},\ }\bibfield  {title} {\bibinfo {title} {Quantum algorithms revisited},\ }\href@noop {} {\bibfield  {journal} {\bibinfo  {journal} {Proceedings of the Royal Society of London. Series A: Mathematical, Physical and Engineering Sciences}\ }\textbf {\bibinfo {volume} {454}},\ \bibinfo {pages} {339} (\bibinfo {year} {1998})}\BibitemShut {NoStop}%
\bibitem [{\citenamefont {Cornelissen}\ \emph {et~al.}(2021)\citenamefont {Cornelissen}, \citenamefont {Mande}, \citenamefont {Ozols},\ and\ \citenamefont {de~Wolf}}]{cornelissen2021exact}%
  \BibitemOpen
  \bibfield  {author} {\bibinfo {author} {\bibfnamefont {A.}~\bibnamefont {Cornelissen}}, \bibinfo {author} {\bibfnamefont {N.~S.}\ \bibnamefont {Mande}}, \bibinfo {author} {\bibfnamefont {M.}~\bibnamefont {Ozols}},\ and\ \bibinfo {author} {\bibfnamefont {R.}~\bibnamefont {de~Wolf}},\ }\bibfield  {title} {\bibinfo {title} {Exact quantum query complexity of computing hamming weight modulo powers of two and three},\ }\href@noop {} {\bibfield  {journal} {\bibinfo  {journal} {arXiv preprint arXiv:2112.14682}\ } (\bibinfo {year} {2021})}\BibitemShut {NoStop}%
\bibitem [{\citenamefont {Ambainis}\ \emph {et~al.}(2017{\natexlab{b}})\citenamefont {Ambainis}, \citenamefont {Iraids},\ and\ \citenamefont {Nagaj}}]{ambainis2017exact}%
  \BibitemOpen
  \bibfield  {author} {\bibinfo {author} {\bibfnamefont {A.}~\bibnamefont {Ambainis}}, \bibinfo {author} {\bibfnamefont {J.}~\bibnamefont {Iraids}},\ and\ \bibinfo {author} {\bibfnamefont {D.}~\bibnamefont {Nagaj}},\ }\bibfield  {title} {\bibinfo {title} {Exact quantum query complexity of $\mathrm{EXACT}_{k,l}^n$},\ }in\ \href@noop {} {\emph {\bibinfo {booktitle} {International Conference on Current Trends in Theory and Practice of Informatics}}}\ (\bibinfo {organization} {Springer},\ \bibinfo {year} {2017})\ pp.\ \bibinfo {pages} {243--255}\BibitemShut {NoStop}%
\bibitem [{\citenamefont {Liu}\ and\ \citenamefont {Nocedal}(1989)}]{liu1989limited}%
  \BibitemOpen
  \bibfield  {author} {\bibinfo {author} {\bibfnamefont {D.~C.}\ \bibnamefont {Liu}}\ and\ \bibinfo {author} {\bibfnamefont {J.}~\bibnamefont {Nocedal}},\ }\bibfield  {title} {\bibinfo {title} {On the limited memory bfgs method for large scale optimization},\ }\href@noop {} {\bibfield  {journal} {\bibinfo  {journal} {Mathematical programming}\ }\textbf {\bibinfo {volume} {45}},\ \bibinfo {pages} {503} (\bibinfo {year} {1989})}\BibitemShut {NoStop}%
\bibitem [{\citenamefont {Paszke}\ \emph {et~al.}(2019)\citenamefont {Paszke}, \citenamefont {Gross}, \citenamefont {Massa}, \citenamefont {Lerer}, \citenamefont {Bradbury}, \citenamefont {Chanan}, \citenamefont {Killeen}, \citenamefont {Lin}, \citenamefont {Gimelshein}, \citenamefont {Antiga}, \citenamefont {Desmaison}, \citenamefont {Köpf}, \citenamefont {Yang}, \citenamefont {DeVito}, \citenamefont {Raison}, \citenamefont {Tejani}, \citenamefont {Chilamkurthy}, \citenamefont {Steiner}, \citenamefont {Fang}, \citenamefont {Bai},\ and\ \citenamefont {Chintala}}]{paszke2019pytorch}%
  \BibitemOpen
  \bibfield  {author} {\bibinfo {author} {\bibfnamefont {A.}~\bibnamefont {Paszke}}, \bibinfo {author} {\bibfnamefont {S.}~\bibnamefont {Gross}}, \bibinfo {author} {\bibfnamefont {F.}~\bibnamefont {Massa}}, \bibinfo {author} {\bibfnamefont {A.}~\bibnamefont {Lerer}}, \bibinfo {author} {\bibfnamefont {J.}~\bibnamefont {Bradbury}}, \bibinfo {author} {\bibfnamefont {G.}~\bibnamefont {Chanan}}, \bibinfo {author} {\bibfnamefont {T.}~\bibnamefont {Killeen}}, \bibinfo {author} {\bibfnamefont {Z.}~\bibnamefont {Lin}}, \bibinfo {author} {\bibfnamefont {N.}~\bibnamefont {Gimelshein}}, \bibinfo {author} {\bibfnamefont {L.}~\bibnamefont {Antiga}}, \bibinfo {author} {\bibfnamefont {A.}~\bibnamefont {Desmaison}}, \bibinfo {author} {\bibfnamefont {A.}~\bibnamefont {Köpf}}, \bibinfo {author} {\bibfnamefont {E.}~\bibnamefont {Yang}}, \bibinfo {author} {\bibfnamefont {Z.}~\bibnamefont {DeVito}}, \bibinfo {author} {\bibfnamefont {M.}~\bibnamefont {Raison}}, \bibinfo {author} {\bibfnamefont {A.}~\bibnamefont {Tejani}}, \bibinfo
  {author} {\bibfnamefont {S.}~\bibnamefont {Chilamkurthy}}, \bibinfo {author} {\bibfnamefont {B.}~\bibnamefont {Steiner}}, \bibinfo {author} {\bibfnamefont {L.}~\bibnamefont {Fang}}, \bibinfo {author} {\bibfnamefont {J.}~\bibnamefont {Bai}},\ and\ \bibinfo {author} {\bibfnamefont {S.}~\bibnamefont {Chintala}},\ }\href@noop {} {\bibinfo {title} {Pytorch: An imperative style, high-performance deep learning library}} (\bibinfo {year} {2019}),\ \Eprint {https://arxiv.org/abs/1912.01703} {arXiv:1912.01703 [cs.LG]} \BibitemShut {NoStop}%
\bibitem [{\citenamefont {Virtanen}\ \emph {et~al.}(2020)\citenamefont {Virtanen}, \citenamefont {Gommers}, \citenamefont {Oliphant}, \citenamefont {Haberland}, \citenamefont {Reddy}, \citenamefont {Cournapeau}, \citenamefont {Burovski}, \citenamefont {Peterson}, \citenamefont {Weckesser}, \citenamefont {Bright}, \citenamefont {{van der Walt}}, \citenamefont {Brett}, \citenamefont {Wilson}, \citenamefont {Millman}, \citenamefont {Mayorov}, \citenamefont {Nelson}, \citenamefont {Jones}, \citenamefont {Kern}, \citenamefont {Larson}, \citenamefont {Carey}, \citenamefont {Polat}, \citenamefont {Feng}, \citenamefont {Moore}, \citenamefont {{VanderPlas}}, \citenamefont {Laxalde}, \citenamefont {Perktold}, \citenamefont {Cimrman}, \citenamefont {Henriksen}, \citenamefont {Quintero}, \citenamefont {Harris}, \citenamefont {Archibald}, \citenamefont {Ribeiro}, \citenamefont {Pedregosa}, \citenamefont {{van Mulbregt}},\ and\ \citenamefont {{SciPy 1.0 Contributors}}}]{2020SciPy-NMeth}%
  \BibitemOpen
  \bibfield  {author} {\bibinfo {author} {\bibfnamefont {P.}~\bibnamefont {Virtanen}}, \bibinfo {author} {\bibfnamefont {R.}~\bibnamefont {Gommers}}, \bibinfo {author} {\bibfnamefont {T.~E.}\ \bibnamefont {Oliphant}}, \bibinfo {author} {\bibfnamefont {M.}~\bibnamefont {Haberland}}, \bibinfo {author} {\bibfnamefont {T.}~\bibnamefont {Reddy}}, \bibinfo {author} {\bibfnamefont {D.}~\bibnamefont {Cournapeau}}, \bibinfo {author} {\bibfnamefont {E.}~\bibnamefont {Burovski}}, \bibinfo {author} {\bibfnamefont {P.}~\bibnamefont {Peterson}}, \bibinfo {author} {\bibfnamefont {W.}~\bibnamefont {Weckesser}}, \bibinfo {author} {\bibfnamefont {J.}~\bibnamefont {Bright}}, \bibinfo {author} {\bibfnamefont {S.~J.}\ \bibnamefont {{van der Walt}}}, \bibinfo {author} {\bibfnamefont {M.}~\bibnamefont {Brett}}, \bibinfo {author} {\bibfnamefont {J.}~\bibnamefont {Wilson}}, \bibinfo {author} {\bibfnamefont {K.~J.}\ \bibnamefont {Millman}}, \bibinfo {author} {\bibfnamefont {N.}~\bibnamefont {Mayorov}}, \bibinfo {author} {\bibfnamefont
  {A.~R.~J.}\ \bibnamefont {Nelson}}, \bibinfo {author} {\bibfnamefont {E.}~\bibnamefont {Jones}}, \bibinfo {author} {\bibfnamefont {R.}~\bibnamefont {Kern}}, \bibinfo {author} {\bibfnamefont {E.}~\bibnamefont {Larson}}, \bibinfo {author} {\bibfnamefont {C.~J.}\ \bibnamefont {Carey}}, \bibinfo {author} {\bibfnamefont {{\.I}.}~\bibnamefont {Polat}}, \bibinfo {author} {\bibfnamefont {Y.}~\bibnamefont {Feng}}, \bibinfo {author} {\bibfnamefont {E.~W.}\ \bibnamefont {Moore}}, \bibinfo {author} {\bibfnamefont {J.}~\bibnamefont {{VanderPlas}}}, \bibinfo {author} {\bibfnamefont {D.}~\bibnamefont {Laxalde}}, \bibinfo {author} {\bibfnamefont {J.}~\bibnamefont {Perktold}}, \bibinfo {author} {\bibfnamefont {R.}~\bibnamefont {Cimrman}}, \bibinfo {author} {\bibfnamefont {I.}~\bibnamefont {Henriksen}}, \bibinfo {author} {\bibfnamefont {E.~A.}\ \bibnamefont {Quintero}}, \bibinfo {author} {\bibfnamefont {C.~R.}\ \bibnamefont {Harris}}, \bibinfo {author} {\bibfnamefont {A.~M.}\ \bibnamefont {Archibald}}, \bibinfo {author}
  {\bibfnamefont {A.~H.}\ \bibnamefont {Ribeiro}}, \bibinfo {author} {\bibfnamefont {F.}~\bibnamefont {Pedregosa}}, \bibinfo {author} {\bibfnamefont {P.}~\bibnamefont {{van Mulbregt}}},\ and\ \bibinfo {author} {\bibnamefont {{SciPy 1.0 Contributors}}},\ }\bibfield  {title} {\bibinfo {title} {{{SciPy} 1.0: Fundamental Algorithms for Scientific Computing in Python}},\ }\href {https://doi.org/10.1038/s41592-019-0686-2} {\bibfield  {journal} {\bibinfo  {journal} {Nature Methods}\ }\textbf {\bibinfo {volume} {17}},\ \bibinfo {pages} {261} (\bibinfo {year} {2020})}\BibitemShut {NoStop}%
\bibitem [{Note1()}]{Note1}%
  \BibitemOpen
  \bibinfo {note} {\protect \url {https://github.com/wuzp15/VarQQA}}\BibitemShut {NoStop}%
\bibitem [{\citenamefont {Ye}(2023)}]{ye2023exact}%
  \BibitemOpen
  \bibfield  {author} {\bibinfo {author} {\bibfnamefont {Z.}~\bibnamefont {Ye}},\ }\href@noop {} {\bibinfo {title} {On the exact quantum query complexity of $\text{MOD}_m^n$ and $\text{EXACT}_{k,l}^n$}} (\bibinfo {year} {2023}),\ \Eprint {https://arxiv.org/abs/2303.10935} {arXiv:2303.10935 [quant-ph]} \BibitemShut {NoStop}%
\bibitem [{\citenamefont {Cayley}(1846)}]{Cayley+1846+119+123}%
  \BibitemOpen
  \bibfield  {author} {\bibinfo {author} {\bibfnamefont {A.}~\bibnamefont {Cayley}},\ }\bibfield  {title} {\bibinfo {title} {Sur quelques propriétés des déterminants gauches.},\ }\href {https://doi.org/doi:10.1515/crll.1846.32.119} {\ \textbf {\bibinfo {volume} {1846}},\ \bibinfo {pages} {119} (\bibinfo {year} {1846})}\BibitemShut {NoStop}%
\bibitem [{\citenamefont {Zhou}\ \emph {et~al.}(2003)\citenamefont {Zhou}, \citenamefont {Zeng}, \citenamefont {Xu},\ and\ \citenamefont {Sun}}]{zhou2003quantum}%
  \BibitemOpen
  \bibfield  {author} {\bibinfo {author} {\bibfnamefont {D.}~\bibnamefont {Zhou}}, \bibinfo {author} {\bibfnamefont {B.}~\bibnamefont {Zeng}}, \bibinfo {author} {\bibfnamefont {Z.}~\bibnamefont {Xu}},\ and\ \bibinfo {author} {\bibfnamefont {C.}~\bibnamefont {Sun}},\ }\bibfield  {title} {\bibinfo {title} {Quantum computation based on d-level cluster state},\ }\href@noop {} {\bibfield  {journal} {\bibinfo  {journal} {Physical Review A}\ }\textbf {\bibinfo {volume} {68}},\ \bibinfo {pages} {062303} (\bibinfo {year} {2003})}\BibitemShut {NoStop}%
\bibitem [{\citenamefont {de~Guise}\ \emph {et~al.}(2018)\citenamefont {de~Guise}, \citenamefont {Di~Matteo},\ and\ \citenamefont {S{\'a}nchez-Soto}}]{de2018simple}%
  \BibitemOpen
  \bibfield  {author} {\bibinfo {author} {\bibfnamefont {H.}~\bibnamefont {de~Guise}}, \bibinfo {author} {\bibfnamefont {O.}~\bibnamefont {Di~Matteo}},\ and\ \bibinfo {author} {\bibfnamefont {L.~L.}\ \bibnamefont {S{\'a}nchez-Soto}},\ }\bibfield  {title} {\bibinfo {title} {Simple factorization of unitary transformations},\ }\href@noop {} {\bibfield  {journal} {\bibinfo  {journal} {Physical Review A}\ }\textbf {\bibinfo {volume} {97}},\ \bibinfo {pages} {022328} (\bibinfo {year} {2018})}\BibitemShut {NoStop}%
\bibitem [{\citenamefont {McClean}\ \emph {et~al.}(2018)\citenamefont {McClean}, \citenamefont {Boixo}, \citenamefont {Smelyanskiy}, \citenamefont {Babbush},\ and\ \citenamefont {Neven}}]{mcclean2018barren}%
  \BibitemOpen
  \bibfield  {author} {\bibinfo {author} {\bibfnamefont {J.~R.}\ \bibnamefont {McClean}}, \bibinfo {author} {\bibfnamefont {S.}~\bibnamefont {Boixo}}, \bibinfo {author} {\bibfnamefont {V.~N.}\ \bibnamefont {Smelyanskiy}}, \bibinfo {author} {\bibfnamefont {R.}~\bibnamefont {Babbush}},\ and\ \bibinfo {author} {\bibfnamefont {H.}~\bibnamefont {Neven}},\ }\bibfield  {title} {\bibinfo {title} {Barren plateaus in quantum neural network training landscapes},\ }\href@noop {} {\bibfield  {journal} {\bibinfo  {journal} {Nature communications}\ }\textbf {\bibinfo {volume} {9}},\ \bibinfo {pages} {4812} (\bibinfo {year} {2018})}\BibitemShut {NoStop}%
\end{thebibliography}%
	
\end{document}